\documentstyle[preprint,eqsecnum,aps]{revtex}

\newcommand{\f}{\varphi}
\newcommand{\G}{G_{\mu \nu}}
\newcommand{\g}{g_{\mu \nu}}
\newcommand{\gu}{g^{\mu \nu}}
\newcommand{\tg}{\tilde{g}_{\mu \nu}}
\newcommand{\F}{\Phi}

\newcommand{\tgu}{\tilde{g}^{\mu \nu}}
\newcommand{\tC}{\tilde{\Lambda}(\F)}
\newcommand{\C}{\Lambda(\f)}

\begin{document}

\draft

\title{Scalar-Tensor Cosmologies and their Late Time Evolution}

\author{ David I. Santiago\thanks{david@spacetime.stanford.edu}}
\address{ Department of Physics, Stanford University }
\author{ Dimitri Kalligas} 
\address{ Physics Department, University of Athens }
\author{ Robert V. Wagoner\thanks{wagoner@leland.stanford.edu}}
\address{ Department of Physics, Stanford University}
\date{\today}

\maketitle

\begin{abstract}
We study the asymptotic behavior at late times of Friedmann-Robertson-Walker 
(uniform density) cosmological models within scalar-tensor theories of gravity.
Particularly, we analyze the late time behavior in the present (matter 
dominated) epoch of the universe. The result of Damour and Nordtvedt that for a
massless scalar in a flat cosmology the Universe evolves towards a state 
indistinguishable from general relativity is generalized. We first study a 
massless scalar field in an open universe. It is found that, while the universe
tends to approach a state with less scalar contribution to gravity, the 
attractor mechanism is not effective enough to drive the theory towards a final
state indistinguishable from general relativity. For the self-interacting case 
it is found that the scalar field potential dominates the late time behavior. 
In most cases this makes the attractor mechanism effective, thus resulting in 
a theory of gravity with vanishingly small scalar contribution even for the 
open Universe. 
\end{abstract}


\section{Introduction}
\label{sec-int}

One can argue that the simplest generalization of Einstein's theory of general 
relativity (GR) is a scalar-tensor (ST) theory of gravity. The simplest and 
earliest scalar-tensor theories\cite{jr,fr,bd} considered a massless scalar 
field with constant coupling to matter. Later, ST gravity was generalized by 
having a scalar field self-interaction and a dynamical coupling to matter\cite
{brg,ndt,wag1}. More recently ST theories of gravity have been generalized
further to the case of multiple scalar fields\cite{dm-ef}. 

Our main aim is to generalize a very important result of Damour and Nordtvedt
\cite{dm-nd}. They consider multiple massless scalar fields with dynamical
coupling to matter. For Friedmann-Robertson-Walker (FRW) cosmological models 
they derive a master equation for the scalar fields. This equation greatly 
simplifies the analysis of the scalar fields' evolution. Using their master 
equation, Damour and Nordtvedt (DN) show that, for flat cosmologies in a matter
dominated epoch, a wide class of ST theories evolve towards a state with no 
scalar admixture to gravity. This means that in such a class of ST models 
their predictions evolve toward those of GR.

We first generalize DN by considering an open cosmological model in ST gravity
with a massless scalar field (for simplicity we consider a single scalar). It 
is found that, in a matter dominated epoch, a wide class of ST theories evolve 
towards a state with less scalar field contribution. In general, the scalar 
field contribution to gravity does not vanish as is the case for a flat 
cosmology. Therefore in an open universe the predictions of our class of ST 
theories will be evolving closer to those of GR, but they will be 
distinguishable from those of GR in the final asymptotic state.

A lot of work has been focused on scalar-tensor theories in which the scalar 
field/fields is/are massless\cite{knw,sr-al,wag-ka,skw}. There are 
arguments as to why a scalar field might be massless\cite{str1,str2}. But, in 
all probability, the scalar field will acquire a self-interaction after quantum
corrections because there is no gauge or symmetry principle to protect it from
developing a potential term in the effective Lagrangian. Therefore if ST 
gravity is the low energy limit of a possible quantum theory of gravity, it
will probably have a self-interacting scalar field. These considerations lead 
us to further generalize DN by including a scalar field potential in our 
Lagrangian for ST gravity. We find that the scalar field evolution is dominated
at large times by the self-interaction term. The presence of the 
self-interaction term (for very general forms of the self-interaction) makes 
the force term in the scalar field evolution equation comparable to the 
friction term for late times. This results in there being a wide classes of ST 
theories whose predictions asymptotically become those of GR for both flat and 
negatively curved cosmologies in the matter era.

We see that, for various cosmologies, wide classes of ST theories of gravity 
approach the predictions of GR in the present matter epoch. The results of 
solar system experiments\cite{will1,vm,vlbi,llr} imply that gravity behaves 
very closely to what GR predicts at the present epoch in the weak field regime.
In general, this doesn't necessarily mean that GR is the correct theory of 
gravity because ST theories can and do evolve closer to GR in the present 
epoch.

Throughout this paper we follow the conventions of MTW\cite{mtw} with $G=c=
1$.

\section{Scalar-Tensor Theories of Gravity}
\label{STG}

We consider the most general scalar-tensor theories of gravity with a single
scalar field. In these theories the gravitational interaction is mediated by
a spin-2 field, the metric $\tg$, and a spin-0 field, a scalar field $\F$.

The field equations for these theories follow from the action\cite{dm-ef,dm-nd}
\begin{eqnarray}
S=\frac{1}{16\pi}\int d^{4}x\sqrt{-\tilde{g}}\,[\F \tilde{R} - 
\frac{\omega(\F)}{\F}\tgu\partial_{\mu}\F\partial_{\nu}\F -2\tC] \nonumber \\
 +\;\; S_{m}[\Psi_{m},\tg]  ,
\end{eqnarray}
where $\tilde{R}$ is the Ricci scalar constructed from the physical metric
$\tg$, $\omega(\F)$ is the coupling function of the scalar field to matter,
and the cosmological term $\tC$ is the scalar field potential. The scalar field
$\F$ plays the role of the inverse gravitational constant $G^{-1}$. The last 
term is the action of the matter fields, $\Psi_{m}$, which couple only to the 
metric $\tg$ and not to the scalar field $\F$ in order to satisfy the weak 
equivalence principle. Each possible specification of the two arbitrary 
functions $\omega(\F)$ and $\tC$ defines a different scalar-tensor theory. In
general, ST gravity are theories of gravity with a varying cosmological
 ``constant'' and varying gravitational ``constant''.

The field equations which follow from this action by varying $\tg$ and $\F$
are quite inconvenient because they mix the spin-2 and spin-0 excitations\cite
{dm-nd}. We are going to define two new field variables, $\g$ and $\f \,$, by a
 conformal transformiation that disentangles the two propagation modes. 
This will be called the spin frame while the old one will be called the 
physical frame. We make the following definitions
\begin{eqnarray}
\tg \equiv A^2(\f)\g \label{eq:pg} \\
\F \equiv \frac{1}{A^2(\f)} \label{eq:Phi} \\
\C \equiv A^4(\f)\tC \label{eq:lamb} \\
\alpha(\f) \equiv \frac{d \ln A(\f)}{d \f} \, . 
\end{eqnarray}
In order to uniquely define the new spin-frame quantities, $\g$, $\f$ and 
$A(\f)$, in terms of the old physical frame quantities, $\tg$, $\F$ and $\omega
(\F)$ we impose the condition
\begin{equation}
\alpha^2(\f) = \frac{1}{2\omega(\F) + 3}. 
\end{equation}
In terms of the spin-frame quantities the action becomes
\begin{eqnarray}
S=\frac{1}{16\pi}\int d^{4}x\sqrt{-g}\,[R - 2\gu\partial_{\mu}\f\partial_{\nu}
\f -2\C] \nonumber \\
 +\;\; S_{m}[\Psi_{m},A^2(\f)\g]  \label{eq:S}.
\end{eqnarray}
Note that the matter fields, $\Psi_{m}$, couple only to the physical metric $
\tg=A^2(\f)\g$. As previously mentioned this preserves the equivalence 
principle. Therefore test particles follow geodesics of the physical metric. In
this frame specification of the arbitrary functions $A(\f)$ and $\Lambda(\f)$ 
uniquely defines a scalar-tensor theory of gravity.

By varying the action with respect to $\g$ and $\f \,$ we obtain the field 
equations
\begin{eqnarray}
\G+\g \C=8\pi T_{\mu \nu} + 2\,(\f_{,\mu} \f_{, \nu} -\frac{1}{2}\g\f^{,\sigma}
\f_{,\sigma}) \label{eq:G} \\
\Box \f -\frac{1}{2}\frac{d\C}{d \f}= -4\pi \alpha(\f)T\, , \label{eq:f}
\end{eqnarray}
where
\begin{eqnarray}
T^{\mu \nu}=\frac{2}{\sqrt{-g}}\frac{\delta S_{m}}{\delta \g} \label{eq:T}\\
T=\gu T_{\mu \nu} \\
\Box \f= \f^{, \mu}\,_{;\mu}
\end{eqnarray}
are the spin-frame energy-momentum tensor, its trace, and the divergence of 
the gradient of $\f$.

It is important to recognize that the spin frame energy-momentum tensor, $
T^{\mu \nu}$, is not the physical energy-momentum tensor, i.e., that measured 
in a local Lorentz frame. The physical energy-momentum is defined by varying 
the action with respect to the physical metric, eq. (\ref{eq:pg}),
\begin{equation}
\tilde{T}^{\mu \nu}=\frac{2}{\sqrt{-\tilde{g}}}\frac{\delta S_{m}}{\delta \tg}
\, .
\end{equation}
It is related to the spin-frame energy-momentum tensor by the relations
\begin{eqnarray}
T_{\mu \nu}=A^{2}(\f)\tilde{T}_{\mu \nu}\, , \\
T_{\mu}\,^{ \nu}=A^{4}(\f)\tilde{T}_{\mu}\,^{ \nu}\, , \\
T^{\mu \nu}=A^{6}(\f)\tilde{T}^{\mu \nu}\, . \label{eq:pt-t}
\end{eqnarray}
The physical energy-momentum tensor is covariantly conserved with respect to 
the physical metric. The spin frame energy-momentum tensor is not conserved 
because it can interchange momentum and energy with the scalar field. The 
Bianchi identity implies that the sum of the energy-momentum tensor of the 
matter, $T^{\mu \nu}$, of the scalar field,
\begin{equation}
S_{\mu \nu}=\frac{1}{4 \pi}\,(\f_{,\mu} \f_{, \nu} -\frac{1}{2}\g\f^{,\sigma}
\f_{,\sigma})\, ,
\end{equation}
and of the cosmological term,
\begin{equation}
-\frac{1}{8 \pi} \g \C\, ,
\end{equation}
is conserved. Using the field equations this conservation law can be written in
the very compact form 
\begin{equation}
T_{\mu}\,^{\nu}\,_{; \nu}= \alpha(\f)T \f_{, \mu}\,, \label{eq:bianchi}
\end{equation}
where a semicolon represents covariant derivative with respect to the spin 
frame metric $\g$.

The physical frame is given its name because matter couples directly only to 
its metric. This is very natural since, because of the direct coupling, 
particles have constant mass and move on geodesics of the physical metric, 
i.e., the physical energy-momentum tensor is conserved. When one considers the 
spin-frame one can think of the particles as having a scalar field dependent 
mass\cite{bd} or, equivalently, having a constant mass but ``feeling'' a scalar
field dependent force. This manifests itself in the conservation of the sum 
of the spin-frame energy-momentum tensors of matter fields, scalar field, and 
cosmological term. Therefore it is really a matter of interpretation which one 
is more fundamental, except that the physical metric is the one that defines
lengths and rates of ideal clocks. We work in the spin-frame representation 
because the equations are in a more manageable form.

\section{Scalar-Tensor FRW Cosmologies}
\label{cosm}

We consider Friedmann-Robertson-Walker (FRW) cosmologies. These describe 
homogeneous and isotropic universes. Since our universe is homogeneous and 
isotropic to a high degree on large scales, a FRW cosmology is a very good 
approximation for the cosmology we live in. Therefore we consider spacetimes 
with a FRW metric
\begin{equation}
ds^{2}=-dt^{2}+R^{2}(t)\,[\frac{dr^{2}}{1-kr^{2}}+r^{2}d\Omega ^{2}]\, ,
\end{equation}
where $d\Omega$ is the element of solid angle and $k=-1,\ 0,\ \mbox{or}\ 1$ 
according to whether the universe is open, flat, or closed. The condition of 
homogeneity implies that our scalar field is only a function of the spin-frame 
time coordinate, $\f=\f(t)$. One must remember that the spin-frame ``proper'' 
time, $d\tau=\sqrt{-ds^2}$, is not the proper time measured by ideal clocks. 
Since particles couple to the physical metric (\ref{eq:pg}), the proper time 
measured by ideal clocks is $d\tilde{\tau}=A(\f)d\tau$. In particular, comoving
observers have proper time $d\tilde{\tau}=A(\f)dt$. Since the physical metric 
also defines lengths, the physical scale factor is given by $\tilde{R}(t) 
\equiv A(\f)R(t)$, where $R(t)$ is the spin-frame scale factor.

The only stress-energy tensor consistent with the assumptions that space is 
homogeneous and isotropic is one that is diagonal and spatially isotropic. 
Therefore matter in a FRW cosmology behaves like a perfect fluid
\begin{equation}
T^{\mu \nu}=(\rho + p)u^{\mu}u^{\nu}+p\gu \, ,
\end{equation}
where
\begin{equation}
u^{\mu}\equiv \frac{dx^{\mu}}{d\tau}
\end{equation}
is the spin-frame 4-velocity normalized to $\g u^{\mu}u^{\nu}=-1$ for massive
particles. Since $d\tilde{\tau}=A(\f)d\tau$, we have $u^{\mu}=A(\f)\tilde{u}^
{\mu}$ with 
\begin{equation}
\tilde{u}^{\mu}\equiv \frac{dx^{\mu}}{d\tilde{\tau}}
\end{equation}
being the physical 4-velocity, i.e., the actual 4-velocity particles have.
The relation (\ref{eq:pt-t}) between the physical and spin-frame 
energy-momentum tensors leads to the following expression for the spin-frame 
density and pressure of the matter in terms of the physical quantities,
\begin{equation}
\rho=A^{4}(\f)\tilde{\rho},\ p=A^{4}(\f)\tilde{p}\, . \label{eq:p}
\end{equation}
For the FRW metric and perfect fluid energy-momentum tensor the field equations
are
\begin{eqnarray}
-3\frac{\ddot{R}}{R}=4\pi (\rho + 3p) + 2(\dot{\f})^{2}-\C\,, \label{eq:acc} \\
3\left( \frac{\dot{R}}{R} \right)^{2} + 3\frac{k}{R^{2}}=8\pi \rho + (\dot{\f})
^{2} + \C\label{eq:h} \\
\ddot{\f} + 3\frac{\dot{R}}{R}\dot{\f}=-4\pi \alpha (\f)(\rho - 3p) -
\frac{1}{2}\frac{d \C}{d \f}\, , \label{eq:sc}
\end{eqnarray}
with Bianchi identity
\begin{equation}
d(\rho R^{3}) + pd(R^{3})=(\rho - 3p)R^{3}da(\f)\, .
\end{equation}

Following Damour and Nordtvedt\cite{dm-nd} closely, we proceed to derive a 
``master'' equation for the scalar field that provides a more transparent
interpretation of the scalar field dynamics. We define
\begin{equation}
\chi=\ln (\frac{R}{\text{constant}}) \, .
\end{equation}
In terms of $\chi$ the field equations and Bianchi identity become
\begin{eqnarray}
-3\ddot{\chi} -3\dot{\chi}^2=2\dot{\chi}^2(\f\prime)^2 + 4\pi(\rho + 3p)
  -\C \label {eq:chi1}\\
3\dot{\chi}^2 +3k e^{-2\chi}=\dot{\chi}^2(\f\prime)^2 + 8\pi \rho +
\C \label{eq:chi2} \\
\dot{\chi}^2 \f\prime\prime + (\ddot{\chi} + 3\dot{\chi}^2)\f\prime=-4\pi
\alpha(\f)(\rho -3p) -\frac{1}{2} \frac{d\C}{d\f} \label{eq:chif} \\
d(\rho e^{3\chi}) + pd(e^{3\chi})=(\rho -3p)e^{3\chi}da(\f) \label{eq:cons}\, ,
\end{eqnarray}
where primes denote derivatives with respect to $\chi$ and $a(\f) \equiv
\ln A(\f)$. We use equations (\ref{eq:chi1}) and (\ref{eq:chi2}) along with the
definitions
\begin{eqnarray}
\epsilon \equiv \frac{3k e^{-2\chi}}{8 \pi \rho} \\
\eta \equiv \frac{p}{\rho} \\
\lambda \equiv \frac{\C}{8 \pi \rho} 
\end{eqnarray}
to rewrite equation (\ref{eq:chif}) in the very useful form
\begin{equation}
\frac{2(1 + \lambda - \epsilon)}{3 - (\f\prime)^2}\f\prime\prime + 
(1 - \eta +2\lambda - \frac{4\epsilon}{3})\f\prime = -\alpha(\f)(1-3\eta)
-\lambda \frac{d\ln \C}{d\f} \,.
\end{equation}
Finally we use equations (\ref{eq:Phi}) and (\ref{eq:lamb}) with the identity
\begin{equation}
\frac{d}{d\f} = -2 \alpha(\f)\F^2\frac{d}{d\F} \, ,
\end{equation}
to obtain
\begin{equation}
\frac{2(1 + \lambda - \epsilon)}{3 - (\f\prime)^2}\f\prime\prime + 
(1 - \eta +2\lambda - \frac{4\epsilon}{3})\f\prime = -\alpha(\f)[1-3\eta
-2\lambda(\F^2\frac{d}{d\F}\ln \tC -2 \F) ]\, . \label{eq:master}
\end{equation}
This is our master equation. When there is no self-interaction of the scalar
field this equation reduces to
\begin{equation}
\frac{2(1 - \epsilon)}{3 - (\f\prime)^2}\f\prime\prime + 
(1 - \eta - \frac{4\epsilon}{3})\f\prime = -\alpha(\f)[1-3\eta]
 \, . \label{eq:master2}
\end{equation}

If we think of $\f$ as the coordinate of some fictious ``particle''\cite
{dm-nd}, and we measure ``time'' with our $\chi$ variable, the  scalar field 
master evolution equation (\ref{eq:master}) represents the damped motion of the
ficticious particle. Consider the first term in the master equation, i.e., the
acceleration term, $(2[1 + \lambda - \epsilon]/[3 - (\f\prime)^2])\f\prime
\prime$. The effective mass can be considered both velocity dependent through 
the term $2/[3 - (\f\prime)^2]$, and explictly $\chi$-time dependent through $
1 + \lambda - \epsilon$. The appearance of $\lambda \equiv \C / 8 \pi\rho$ in 
the mass term is interesting because one could not have guessed this term by 
imposing in the $\lambda=0$ equation (\ref{eq:master2}) the usual equation of 
state, $p=-\rho$ or $\eta=-1$, for the cosmological term $\C$. We see that, 
although the self-interaction of the scalar field acts as the energy density of
the vacuum, this is not equivalent to $p=-\rho$ in ST gravity. The next term in
the master equation, $(1 - \eta + 2 \lambda - 4 \epsilon / 3)\f\prime$, 
corresponds to a frictional (damping) force linear in the velocity with a $\chi
$-time dependent friction coefficient, $1 - \eta +2\lambda - 4\epsilon / 3$. 
The right hand side of the equation, $-\alpha(\f)[1-3\eta -2\lambda(\F^2
\frac{d}{d\F}\ln \tC -2 \F)]$, is the force term. This force is proportional to
the negative gradient of a potential, $a(\f) \equiv \ln A(\f)$, but the factor 
of proportionality, $1-3\eta -2\lambda(\F^2\frac{d}{d\F}\ln \tC -2 \F)$, is $
\chi$-time dependent. The dynamical system described by the masterequation 
(\ref{eq:master}) is not conservative. This is what one expects because the 
scalar field exchanges energy with the metric and matter fields.

\section{Scalar Field Evolution in FRW Cosmologies}
\label{evol}

In this section we study the scalar field master equation in order to describe 
the cosmological evolution of the scalar field and of its coupling function, $
\alpha(\f)$. We will concentrate on the evolution in the present matter 
dominated (or curvature dominated for sufficiently open universes) epoch. 

Of particular importance will be the asymptotic late time behavior of the 
square of the scalar field coupling function, $\alpha^2(\f)$, because this acts
as a measure of how much ST gravity differs from GR\cite{dm-ef,dm-nd}. The 
closer $\alpha(\f)$ is to zero, the less scalar admixture to gravity there is 
in our ST theory, i. e., the closer its behavior is to the one predicted by GR.
If we take a look at the force term of the master equation (\ref{eq:master}),
\begin{equation}
-\alpha(\f)[1-3\eta-2\lambda(\F^2\frac{d}{d\F}\ln \tC -2 \F)]\, ,
\end{equation}
we see that values of $\f$ that make $\alpha(\f)=0$ are equilibrium points of
the scalar field evolution equation. Whether these equilibrium points are
stable or unstable depends on the sign ($+$ or $-$) of the factor that 
multiplies $\alpha(\f)$ in the force term, on the sign of the mass term, and on
the sign of the friction coefficient in the master evolution equation. All 
three of these terms are $\chi$-time dependent and, therefore, could change 
signs at different values of $\chi$. We now turn to the study of the stable 
equilibrium points (if any exist) of the master equation and the scalar field 
evolution in different cosmological epochs.

\subsection{Radiation Dominated Epoch}
\label{rad}

Throughout most of the radiation era the universe is mostly composed of 
radiation in equilibrium plus a small nonrelativistic matter part:
\begin{eqnarray}
\rho=\rho_{r} + \rho_{m}\\
p=p_{r}=\frac{\rho_{r}}{3}\, .
\end{eqnarray}
The subscripts $r$ and $m$ refer to radiation and nonrelativistic matter 
respectively. For most of the radiation era we have $\rho_{m} \ll \rho_{r}$. 
This last inequality only breaks down near the end of the radiation epoch, but 
only for a duration that is very small compared to the total duration of the 
radiation era. Therefore we only introduce a small error by supposing this 
inequality is true throughout the radiation era.

From baryon conservation and the conservation equation (\ref{eq:cons}) we 
obtain 
\begin{eqnarray}
d(\rho_{m} e^{3\chi})=\rho_{m} e^{3\chi}da(\f) \\
d(\rho_{r} e^{3\chi}) + \frac{\rho_{r}}{3} d(e^{3\chi})=0
\end{eqnarray}
for the evolution of matter and radiation respectively. These equations are
readily integrated to give
\begin{eqnarray}
\rho_{m}=M_{o} e^{-3\chi} e^{a(\f)}=M_{o} A(\f) e^{-3\chi} \label{eq:rhom} \\
\rho_{r}= N_{o} e^{-4\chi} \label{eq:rhor} \, ,
\end{eqnarray}
where $M_{o}$ and $N_{o}$ are constants of integration related to the present
values of the density of matter and radiation.

We will mostly be concerned with open and flat universes (a few words on closed
cosmologies will be said later). For open cosmologies in the radiation epoch it
is a very good approximation to neglect the effect of curvature compared to the
total density and the cosmolgical term, while for flat cosmologies the 
curvature term is nonexistent. Therefore we take
\begin{equation}
\epsilon=0 \, ,
\end{equation}
except for open cosmologies in the matter epoch. Also,
\begin{equation}
\eta \simeq \frac{1}{3} 
\end{equation}
to zeroth order in $\rho_{m} / \rho_{r}$. It is a very good approximation to 
work to zeroth order because of the smallness of $\rho_{m} / \rho_{r}$ through 
most of the radiation epoch. For the cosmological term we have
\begin{equation}
\lambda = \frac{\tC e^{4\chi}}{8 \pi N_{o} \Phi^{2}} =
\frac{\C e^{4\chi}}{8 \pi N_{o}}\,.
\end{equation}
The master scalar field evolution equation (\ref{eq:master}) becomes
\begin{equation}
\frac{\left[1 + \tC e^{4\chi}/(8 \pi N_{o}\Phi^2)\right]}{3 - (\f\prime)^2}
\f\prime\prime + \left(\frac{2}{3} +\frac{\tC e^{4\chi}}{4 \pi N_{o}
\Phi^2} \right)\f\prime =-\frac{e^{4\chi}}{8 \pi N_{o}}\frac{d \C}{d\f}\, .
\end{equation}
The equilibrium points are values of $\f$ that minimize $\C$ if any of them 
exist. In general our scalar field will evolve towards those values during the 
radiation era. We see that $\f$ rolls down the  $\chi$-time dependent potential
$e^{4\chi} \C /(8 \pi N_{o})$. Since $\C = \tC A^{4}(\f)$, $d\C / d\f = 2 
\alpha(\f)[2 \C -(1/ \F) (d \tC / d \F)]$. Therefore the equilibrium points
of $\f$ are either values that make $\alpha(\f)=0$ or values that make $2 \C -
(1/ \F) (d \tC / d \F)=0$. As $\f$ approaches a value that extremizes $A(\f)$, 
$|\alpha(\f)|$ decreases (the extremum condition is $\alpha(\f)=0$), and our ST
gravity behaves more like GR. Therefore in the radiation era our theory can be 
attracted to GR. On the other hand, ST gravity could be attracted to some other
value of the scalar field that minimizes $\C$ and in which the theory does not 
exhibit general relativistic behavior. That our theory can evolve towards GR in
the radiation epoch is a prediction only when there is a cosmological term 
$\C$. In the case when $d \C /d \f$ is zero, the scalar field velocity 
decreases because of friction and it is not attracted to any equilibrium point 
because the force term is zero\cite{dm-nd}.

\subsection{Flat Universe in the Present Epoch}
\label{flat}

The radiation epoch ends when $\rho_{r} \simeq \rho_{m}$. For later times the 
energy density of ordinary matter in the universe is dominated by the
nonrelativistic particles and we are in a matter dominated epoch. The present 
epoch in our universe corresponds to a matter dominated era. If the vacuum 
energy density, i.e., the scalar field self-interaction term, is nonzero, it 
will eventually overtake the matter density and become dominant. In this 
subsection we consider the present epoch in a flat cosmology.

In the present epoch the energy density of radiation is very small compared to
the matter density. Therefore it is a very good approximation to neglect
$\rho_{r}$. The matter is pressureless to a high degree. 
We then have
\begin{equation}
\epsilon=\eta =0 \, ,
\end{equation}
\begin{equation}
\lambda = \frac{\tC e^{3\chi}}{8 \pi M_{o} A(\f) \Phi^{2}} =
\frac{\tC e^{3\chi}}{8 \pi M_{o} \Phi^{3/2}}\,.
\end{equation}
The scalar field master equation (\ref{eq:master}) becomes
\begin{equation}
\frac{2[1 + \tC e^{3\chi}/(8 \pi M_{o} \Phi^{3/2})]}{3 - (\f\prime)^2}
\f\prime\prime + (1 + \frac{\tC e^{3\chi}}{4 \pi M_{o} \Phi^{3/2}})\f\prime = 
-\alpha(\f)[1 -\frac{\tC e^{3\chi}}{4 \pi M_{o} \Phi^{3/2}}(\F^2\frac{d}{d\F}
\ln \tC -2 \F) ] \label{eq:masterflat}\, .
\end{equation}

First consider the case with no cosmological term, $\tC=0$. This was treated by
Damour and Nordtvedt\cite{dm-nd} and we review their results here. For this 
case we have the master equation (\ref{eq:master2}):
\begin{equation}
\frac{2}{3 - (\f\prime)^2}\f\prime\prime + \f\prime = -\alpha(\f) \, .
\end{equation}
The stable equilibrium points are values of the scalar field  that minimize $A
(\f)$, assuming these exist. As $\chi \rightarrow \infty$,  $\f$ will lose 
energy because of friction and will eventually settle around one of its stable
equilibrium points reaching it on the limit. As $\f$ approaches its equilibrium
values, $\alpha(\f)$ approches zero, and ST gravity approches GR. Therefore
when $A(\f)$ has at least a minimum or approches one asymptotically (i. e. $
\alpha(\f) \rightarrow 0$ when $\f \rightarrow \pm \infty$), ST gravity 
exhibits an attractor mechanism towards GR and becomes indistinguishable from 
GR in the limit of large times.

We now consider the special case $\tC = \Phi^2$. The scalar field evolution
equation (\ref{eq:masterflat}) becomes, after dividing by $[1 + \Phi^{1/2} e^{3
\chi} /( 8 \pi M_{o})]$
\begin{equation}
\frac{2}{3 - (\f\prime)^2}
\f\prime\prime + \frac{(8 \pi M_{o} + 2\Phi^{1/2} e^{3\chi})}
{(8 \pi M_{o} + \Phi^{1/2} e^{3\chi})}\f\prime = 
-\frac{\alpha(\f)}{[1 + \Phi^{1/2} e^{3\chi}/(8 \pi M_{o})]} \, .
\end{equation}
The stable equilibrium points are again those values of $\f$ that minimize $A(
\f)$, when any exist. The force term this equation is equal to $-\frac{d}{d\f}
[\ln A(\f)]/[1 + \Phi^{1/2} e^{3\chi} /(8 \pi M_{o})]$. This is a $\chi$-time 
dependent term times the $\f$-gradient of the ``potential'' $\ln A(\f)$. 
Physical lengths and clock times are proportional to the coupling function $A(
\f)=1 / \F^{1/2}$. Therefore it is a reasonable restriction on the possible 
coupling functions that they be finite and nonzero. It is then interesting that
as $\chi \rightarrow \infty$ the friction coefficient approaches 2 while the 
potential, and hence the force term, goes to zero. This implies that the scalar
field will not reach the value that minimizes $A(\f)$ although as $\chi$ 
increases, $\alpha(\f)$ decreases. The attractor mechanism is not effective 
because at large times the potential ``flattens'' while the friction term  
``survives''. We see that in this case ST gravity evolves closer to GR, but it 
does not have GR as a limit for large times. This is an interesting new effect 
in ST gravity. As will be seen later, open universes with no cosmological term 
exhibit the same type of behavior.

We now turn to the case when $\tC \neq \Phi^2$. We first consider the case in 
which $\tC / \F^{3/2}$ does not vanish faster than $e ^{-3\chi}$ as $\chi 
\rightarrow \infty$. Examining our master equation (\ref{eq:masterflat}), we 
see that values of the scalar field that make $\alpha(\f)$ vanish, i.e., values
that minimize or maximize $A(\f)$, are the possible equilibrium points if any 
exist. As $\chi \rightarrow \infty$, the term $-\frac{\tC e^{3\chi}}{4 \pi 
M_{o} \Phi^{3/2}}(\F^2\frac{d}{d\F}\ln \tC -2 \F)$ dominates the force term. 
Therefore if $(\F^2\frac{d}{d\F}\ln \tC -2 \F) \leq 0$ in the neighborhood of 
the equilibrium value of $\f$, values of the scalar field that minimize $A(\f)$
are the stable equilibrium points, otherwise values that maximize $A(\f)$ are 
the stable equilibrium points. The important thing is that in both cases $
\alpha(\f)=0$ at the stable equilibrium point and ST gravity becomes GR at such
a point. As $\chi \rightarrow \infty$, $\f$ evolves towards one of its stable 
equilibrium values (if any exist) and oscillates around it, finally settling at
this equilibrium point in the limit of large $\chi$. If $\tC / \F^{3/2}$ 
vanishes faster than $e ^{-3\chi}$ as $\chi\rightarrow \infty$, we have the 
same situation except that the force term equals $-\alpha(\f)$ asymptotically 
and the equilibrium points are minima of $A(\f)$. Therefore, for $\tC \neq 
\Phi^2$, ST gravity exhibits effective attractor mechanisms and becomes 
indistinguishable from GR in the limit of arbitrarily large times.

\subsection{Open Universe in the Present Epoch}
\label{open}

We now turn to the study of open cosmologies ($k=-1$) in the present epoch.
At the begining of this era, the energy density of the universe is dominated
by the matter density. As time goes on the curvature term becomes dominant and
finally, the vacuum energy density, $\tC$, becomes dominant if it is nonzero.

For an open universe in the present epoch we have
\begin{eqnarray}
\epsilon= -\frac{3 e^{\chi}}{8 \pi M_{0} A(\f)}=  -\frac{3 \Phi^{1/2}e^{\chi}}
{8 \pi M_{0}} \ \ \;\\
\eta =0 \ \ \ \ \ \ \ \ \ \ \ \ \ \ \ \ \ \ \ \ \ \ \ \ \ \ \ \ \ \ \ \ \ \ \
\\
\lambda = \frac{\tC e^{3\chi}}{8 \pi M_{o} A(\f) \Phi^{2}} =
\frac{\tC e^{3\chi}}{8 \pi M_{o} \Phi^{3/2}}\,.
\end{eqnarray}
Our master equation (\ref{eq:master})
becomes
\begin{eqnarray}
\frac{2[1 + \tC e^{3\chi} / (8 \pi M_{o} \Phi^{3/2}) + 3 \Phi^{1/2} e^{\chi} /
(8 \pi M_{0})]}{3 - (\f\prime)^2}\f\prime\prime + (1 + \frac{\tC e^{3\chi}}{4 
\pi M_{o} \Phi^{3/2}} + \frac{ \Phi^{1/2}e^{\chi}}
{2 \pi M_{0}})\f\prime = \nonumber\\
-\alpha(\f)[1 - \frac{\tC e^{3\chi}}{4 \pi M_{o} \Phi^{3/2}}(\F^2\frac{d}{d\F}
\ln \tC -2 \F) ]\, . \label{eq:masteropen}
\end{eqnarray}

First we consider a cosmology with no vacuum energy density, $\tC=0$. The
scalar field evolution equation (\ref{eq:masteropen}) is, upon dividing by $[1 
+ 3 \Phi^{1/2}e^{\chi} (8 \pi M_{0})]$,
\begin{equation}
\frac{2}{3 - (\f\prime)^2}\f\prime\prime + \frac{(8 \pi M_{0} + 4 \Phi^{1/2}
e^{\chi})}{(8 \pi M_{0} + 3 \Phi^{1/2}e^{\chi})}\f\prime = 
-\frac{\alpha(\f)}{[1 + 3 \Phi^{1/2}e^{\chi}/ (8 \pi M_{0})]}\, . 
\end{equation}
Here the stable equilibrium points are values of the scalar field that minimize
$A(\f)$ when they exist. As $\chi \rightarrow \infty$ the friction coefficient 
tends to $4/3$ while the force term vanishes. Therefore although $\f$ will be 
evolving towards one of its stable equilibrium values it will never settle in 
it because the potential ``flattens'', but the friction term 
``survives'' in the limit of arbitrarily large $\chi$. The attractor mechanism 
is not fully effective and ST gravity does not end up in a state 
indistinguishable from GR.

Consider the special case in which $\tC = \Phi^2$. The master equation is
\begin{equation}
\frac{2[1 + \tC e^{3\chi} / (8 \pi M_{o} \Phi^{3/2}) + 3 \Phi^{1/2} e^{\chi} /
(8 \pi M_{0})]}{3 - (\f\prime)^2}\f\prime\prime + (1 + \frac{\Phi^{1/2
} e^{3\chi}}{4 \pi M_{o}} + \frac{ \Phi^{1/2}e^{\chi}}{2 \pi M_{0}})\f\prime =
-\alpha(\f) \,.
\end{equation}
It is obvious that this system will exhibit the same type of behavior as the 
previous case. Therefore we again have an ineffective attractor mechanism.

Now we turn to the more general case $\tC \neq \Phi^2$. After dividing the 
scalar field master equation (\ref{eq:masteropen}) by $[1 + \tC e^{3\chi} / (8 
\pi M_{o} \Phi^{3/2}) + 3 \Phi^{1/2} e^{\chi} /(8 \pi M_{0})]$ we have
\begin{eqnarray}
\frac{2}{3 - (\f\prime)^2}\f\prime\prime + \frac{(8 \pi M_{o} \Phi^{3/2} + 2
\tC e^{3\chi} + 4 \Phi^{2} e^{\chi})}{(8 \pi M_{o} \Phi^{3/2}
 + \tC e^{3\chi} + 3 \Phi^{2} e^{\chi})}\f\prime = \nonumber\\
-\alpha(\f)\frac{[8 \pi M_{o} \Phi^{3/2} - 2 \tC e^{3\chi}(\F^2
\frac{d}{d\F}\ln \tC -2 \F)]}{(8 \pi M_{o} \Phi^{3/2} + \tC e^{3\chi} 
+ 3 \Phi^{2}e^{\chi})}\, . 
\label{eq:masteropen1}
\end{eqnarray}
One can see that if the term $\Phi^{2}e^{\chi}$ is dominant over the term $
\tC e^{3\chi}$ for large times, the potential ``flattens''. Thus the attractor 
mechanism is ineffective and ST gravity does not have GR as a limit for large 
times. If the term $\tC e^{3\chi}$ is dominant, the attractor mechanisn is 
effective and ST gravity converges towards GR in the limit of large times.

\subsection{Simple Examples}

In this section we will consider two representative examples. The first one 
corresponds to an open universe with $\tC =0$, and the second one is a universe
with $\tC \propto \Phi^2$. With these examples we want to show more explicitly 
the main new effect found in the present paper: the attractor mechanism towards
GR exihibited by ST gravity is ineffective in some cases. 

Both of the cases correspond to $\C$ being independent of $\f$. Hence our field
equations are
\begin{eqnarray}
3\left( \frac{\dot{R}}{R} \right)^{2} + 3\frac{k}{R^{2}}=8\pi \rho + (\dot{\f})
^{2} + \Lambda \label{eq:hubble} \\
\ddot{\f} + 3\frac{\dot{R}}{R}\dot{\f}=-4\pi \alpha (\f)(\rho - 3p) \, , \\
d(\rho R^{3}) + pd(R^{3})=(\rho - 3p)R^{3}da(\f)\, .
\end{eqnarray}
In the first case $k=-1$, $\Lambda=0$. In the second case $k=0$ or $-1$, and $
\Lambda$ is a constant different from zero.

Early on the universe is dominated by relativistic particles, i.e., $p=\rho/3$.
Therefore the scalar field equation for both of our cases is
\begin{equation}
\ddot{\f} + 3\frac{\dot{R}}{R}\dot{\f}=0,
\end{equation}
which has the first integral $\dot{\f} \propto R^{-3}$. 
Suppose the kinetic energy density of the scalar field, $\dot{\f}^{2}$ is large
very early on. Since it is proportional to $R^{-6}$, it will be negligible by
the time the radiation epoch ends. Therefore, to first approximation, we can 
neglect the term $\dot{\f}^{2}$ in the field equation (\ref{eq:hubble}) for the
present epoch.

In order to solve our system of equations we need to choose $A(\f)$. Let
$A(\f)=1 + q \f^2 /2$ with q being a positive constant. For this function $\f
=0$ corresponds to $A$ being a minimum at which point the theory will be 
indistinguishable from GR.

First we look at $k=-1$, $\Lambda=0$. We are interested at the behavior for 
large times. In the limit of large times the curvature term is dominant over 
the density of matter. Our field equations become
\begin{eqnarray}
3\left( \frac{\dot{R}}{R} \right)^{2} =\frac{3}{R^2} \\
\ddot{\f} + 3\frac{\dot{R}}{R}\dot{\f}=-4\pi \alpha (\f) \rho  
\label{eq:sc2} \, , \\
d(\rho R^{3})=\rho R^{3}da(\f)\, ,
\end{eqnarray}
because $p \simeq 0$ in the present epoch. The first and third of these 
equations have solutions
\begin{eqnarray}
R=t \ \ \ \ \ \ \ \ \ \ \ \ \ \ \ \ \ \ \ \ \ \ \ \ \ \ \ \ \ \ \ \ \ \ \ \ \\
\rho=\frac{A(\f)M_{o}}{R^3}=\frac{(1 + q \f^2 /2)M_{o}}{t^3}\, ,
\end{eqnarray}
where $M_{o}$ is an integration constant. Using these two the scalar field 
equation (\ref{eq:sc2}) is
\begin{equation}
\ddot{\f} + \frac{3}{t}\dot{\f}=-4 \pi \frac{M_{o}}{t^3} q \f \, .
\end{equation}
The solution to this equation is\cite{math}
\begin{equation}
\f=\frac{1}{t}\left[ c_{1}J_{2}(4\sqrt{\frac{\pi M_{o} q}{t}})
+ c_{2} N_{2}(4\sqrt{\frac{\pi M_{o} q}{t}}) \right] \, ,
\end{equation}
where $J_{2}(x)$ and $N_{2}(x)$ are Bessel functions of the first and second 
kind, and $c_{i}$ with $i=1, \ 2$ are constants of integration determined from 
matching to earlier epochs. As $t \rightarrow \infty$ we have $J_{2}(4\sqrt{
\frac{\pi M_{o} q}{t}}) \rightarrow (2\pi M_{o} q)/t$ and  $N_{2}(4\sqrt{
\frac{\pi M_{o} q}{t}}) \rightarrow -t/ (4 \pi^2 M_{o} q)$; therefore $\f 
\rightarrow -c_{2}/ (4 \pi^2 M_{o} q)$.
We see that asymptotically $\f$ tends to a constant different from zero, but 
thus zero is the value that makes the theory indistinguishable from GR. The 
attractor mechanism is ineffective.

Now we turn to the case where $\Lambda$ is a constant different from zero. 
For large times, $\Lambda$ is dominant over the curvature and matter density 
terms. Our field equations are
\begin{eqnarray}
3\left( \frac{\dot{R}}{R} \right)^{2} = \Lambda \, ,\\
\ddot{\f} + 3\frac{\dot{R}}{R}\dot{\f}=-4\pi \alpha (\f) \rho  \, , \\
d(\rho R^{3})=\rho R^{3}da(\f)\, ,
\end{eqnarray}
The first and third equation have solutions
\begin{eqnarray}
R=L \exp(t\sqrt{\Lambda/3}) \, ,\ \ \ \ \ \ \ \ \ \ \ \ \ \ \ \ \ \ \ \ \ \ 
 \ \ \ \ \ \ \ \ \ \ \ \\
\rho=\frac{A(\f)M_{o}}{R^3}=\frac{(1 + q \f^2 /2)M_{o}}{L^3}\exp(-t\sqrt{3
\Lambda})\, ,
\end{eqnarray}
where $L$ is a constant of integration. The remaining equation is
\begin{equation}
\ddot{\f} + \sqrt{3\Lambda}\dot{\f}=-4 \pi \frac{M_{o}}{L^3} q \f \exp(-t
\sqrt{3\Lambda})\, .
\end{equation}
When we change independent variables by $x \equiv \exp(-t\sqrt{3\Lambda}) 
/\sqrt{3 \Lambda}$, our equation becomes
\begin{equation}
\frac{d^2 \f}{dx^2} + \frac{ 4 \pi M_{o} q}{\sqrt{3\Lambda} L^3 x} \f = 0 \,
\end{equation}
which has solution \cite{math}
\begin{equation}
\f= c_{1}x^{1/2}J_{1}\left[ \left( \frac{4 \pi M_{o} q x}{\sqrt{3 \Lambda} L^3}
\right)^{1/2}  \right] + c_{2}x^{1/2}N_{1}\left[ \left( \frac{4 \pi M_{o} q 
x}{\sqrt{3 \Lambda} L^3} \right)^{1/2} \right]\, ,
\end{equation}
where $J_{1}(x)$ and $N_{1}(x)$ are Bessel functions of the first and second 
kind, and $c_{i}$ with $i=1,\ 2$ are constants of integration determined from
matching with previous epochs. As $t \rightarrow \infty$, $x \rightarrow 0$, 
we have $J_{1}\left[ \left( \frac{4 \pi M_{o} q x}{\sqrt{3 \Lambda} L^3}
\right)^{1/2}  \right] \rightarrow \left(\frac{ \pi M_{o} q x}{\sqrt{3 \Lambda}
L^3} \right)^{1/2}$ and $ N_{1}\left[ \left( \frac{4 \pi M_{o} q x}{\sqrt{3 
\Lambda} L^3} \right)^{1/2} \right] \rightarrow -\left(\frac{1}{\pi}\frac{
\sqrt{3 \Lambda} L^3}{ \pi M_{o} q x} \right)^{1/2}$. Therefore $\f \rightarrow
-c_{2} \frac{1}{\pi}\left( \frac{\sqrt{3 \Lambda} L^3}{ \pi M_{o} q } \right)^{
1/2}$. We once again see that $\f$ does not have zero as its asymptotic value 
and zero is the value that makes our ST gravity indistinguishable from GR. Thus
again the attractor mechanism is ineffective.

\subsection{Closed Cosmologies}

In a closed ($k=1$) universe one expects the scalar field to evolve in an 
analogous way as in an open universe only in the expanding phase. So as the 
universe expands the scalar field will be evolving towards one of its stable 
equilibrium values while it loses energy through the friction term. ST gravity
will never become indistinguishable from GR because once the universe begins to
contract the friction term will change sign at some point in this phase. 
Therefore the friction becomes pumping, preventing the scalar field from 
settling at one of its equilibrium points and making it climb its potential and
exhibiting behavior that is increasingly different than what is predicted by
GR. Thus the contraction of the universe pumps energy into our scalar field, 
making it climb its potential. In general in a closed cosmology there will be 
an attractor mechanism towards GR in the expanding phase, but in the 
contracting phase ST gravity will move away from GR.

\section{Conclusions}

The main purpose of the present paper was to generalize the result of Damour 
and Nordtvedt~\cite{dm-nd} that ST gravity naturally evolves towards GR in the 
present matter epoch. Their result was valid for a massless scalar field 
and for a flat universe. We eliminated these conditions by considering the 
scalar field to be self-interacting and by considering curved cosmologies.

First, a self-interaction, $\tC$, for the scalar field was introduced. This 
inmediatly produced the effect that ST gravity could be attracted towards GR in
the radiation era . This effect does not occur for a massless scalar field. 
With a flat cosmology in the matter epoch, it was found that ST gravity evolves
towards GR at asymptotically large times except for one choice of $\tC$. For 
$\tC=\F^2$ the attractor mechanism is ineffective and the final asymptotic 
state is distinguishable from GR. Next we considered cosmologies with curved 
comoving spatial sections. Since closed cosmologies do not exhibit an 
attractor mechanism, we focused on open cosmologies in the present epoch. It 
was found that with a massless scalar field these exhibit an ineffective 
attractor mechanism with a final asymptotic state different from GR. When the 
scalar field was considered to be self-interacting the attractor mechanism was 
ineffective for $\tC=\F^2$, and for general $\tC$ when the curvature term 
dominates over the self-interacting term for large times. When the 
self-interacting term dominates over the curavture term at large times the 
attractor mechanism is effective and ST gravity becomes indistinguishable
from GR for arbitrarily large times. It is important to note that, even when 
the attractor meshanism is not effective, ST gravity will start converging 
towards GR for suffiently large times. However, in the limit of 
arbitrarily large times, there will be finite though small differences from 
general relativistic behavior. The fact that in these cases the attractor 
mechanisn is not completely effective is somewhat unexpected effect in ST 
gravity.

We must conclude that (given that sufficient time has elapsed) we expect the 
universe to exhibit behavior that is close to what is predicted by GR, and for 
a fairly broad class of ST theories GR will be the theory of gravity 
asymptotically. Therefore the fact that present day experiments in the 
weak-field regime are very consistent with GR is not surprising because ST 
gravity is expected to naturally evolve closer to GR by the present time. In 
order to search for signatures of a scalar componenet of gravity one must probe
earlier epochs in the evolution of the universe, very relativistic objects, and
gravitational radiation.

\end{document}